\documentclass[12pt, preprint]{aastex}
\begin{document}
\title{THE EFFECT OF MODIFIED GRAVITY ON THE ODDS OF THE BOUND VIOLATIONS OF THE TURN-AROUND RADII}
\author{Jounghun Lee\altaffilmark{1} and 
Baojiu Li\altaffilmark{2}}
\altaffiltext{1}{Astronomy Program, Department of Physics and Astronomy, Seoul National University, 
Seoul 08826, Korea}
\email{jounghun@astro.snu.ac.kr}
\altaffiltext{2}{Institute for Computational Cosmology, Department of Physics, 
Durham University, Durham DH1 3LE, UK}
\begin{abstract}
The turn-around radii of the galaxy groups show the imprint of a long battle between their self-gravitational forces and the accelerating 
space. The standard $\Lambda$CDM cosmology based on the general relativity (GR) predicts the existence of an upper bound on the 
expectation value of the turn-around radius which is rarely violated by individual galaxy groups.  We speculate that a deviation of 
the gravitational law from GR on the cosmological scale could cause an appreciable shift of the mean turn-around radius to higher 
values and make the occurrence of the bound violation more probable.  Analyzing the data from high-resolution N-body simulations for two 
specific models with modified gravity (MG) and the standard GR+$\Lambda$CDM cosmology, we determine the turn-around radii of 
the massive Rockstar groups from the peculiar motions of the galactic halos located in the bound zone where the fifth force generated by 
MG is expected to be at most partially shielded.  We detect a $4\sigma$ signal of difference in the odds of the bound violations between a 
fiducial MG and the GR models, proving that the odds of the bound violations increase with the strength of the fifth force produced by the 
presence of MG.  The advantage of using the odds of the bound violations as a complementary diagnostics to probe the nature of gravity 
is discussed.
\end{abstract}
\section{INTRODUCTION}
\label{sec:intro}

Modified gravity (MG) models presume that the true law of gravity deviates from the general relativity (GR) on the cosmological scale 
and claim that the apparent acceleration of the universe in the present epoch can be explained as a function of MG without resorting to 
anti-gravitational dark energy \citep[see][for a review]{mg_review}. Despite that not even an weak evidence for a failure of GR on the 
cosmological scale has so far been found \citep[e.g.,][]{gr_test10,gr_test11,gr_test_cluster,gr_test12,fr_fromcl15,fr_fromwl16}, 
an observational test of gravity is currently and will be persistently one of the most fundamental topics in cosmology until the origin 
of the cosmic acceleration is {\it physically} understood. 
A variety of diagnostics has been developed not only to detect, if any, the presence of MG \citep[see][for a review]{koyama16} but 
also to break the degeneracy between the MG and the other dark energy (DE) models alternative to the cosmological constant 
($\Lambda$), the most prevalent candidate for DE. 

The dynamic masses of galaxy groups \footnote{Conventionally, a galaxy cluster is defined as a bound object composed more than 
$1000$ galaxies while a galaxy group is less massive object having less than $1000$ galaxies \citep{pad93}. 
As pointed out by \citet{tully15}, however, there is no clear boundary that separates the galaxy groups from the galaxy clusters. 
Following \citet{tully15}, we call both the galaxy clusters and groups the "massive groups" throughout this paper.} provide one of those 
recently developed diagnostics, which have been in the limelight of extensive studies \citep{zhao-etal11,lam-etal12,zu-etal14} 
because of its power to probe the nature of gravity in the local Universe.
MG models are classified by the factors additionally introduced to modify GR such as an extra degree of freedom, higher dimensional 
spacetime, non-locality, higher derivatives, most of which lead to an effective enhancement of gravity (so called, the fifth force) on the 
cosmological scale. The survival of such MG models against the stringent solar system test \citep[e.g., see][and references therein]{will14} 
is deliberately implemented by its screening process through which GR can be restored on the small scale \citep{screen_mg}. 
In the presence of unscreened MG, the dynamic mass of a galaxy group would appear to be higher than its lensing mass since 
the latter depends only on the curvature of space around the group. Thus, any discrepancy between the dynamic and the lensing masses 
of the galaxy groups should indicate the presence of MG and thus can be used to constrain the strength of its consequential fifth force 
\citep{zhao-etal11,zu-etal14}. 

The dynamic mass of a group was conventionally estimated by measuring the velocity dispersions of the luminous central galaxies. 
This conventional estimate, however, would fail to discriminate the dynamic mass from the lensing mass even in the presence of MG 
since GR should be almost completely restored at the locations of the luminous central galaxies. The infall velocities of the satellite 
galaxies located outside the virial radii of the galaxy groups have been suggested as better indicators of the presence of unscreened 
MG. However, it has been concerned that the dependence of the infall velocities of the satellites on the baryonic processes as well as the 
large uncertainties associated with their measurements would contaminate a signal, even if detected, of the difference between the 
dynamic and the lensing masses \citep{lam-etal12,zu-etal14}.

Looking to other dynamical properties of the galaxy groups than their dynamic masses may be necessary to complement the existing 
local probes of gravity on the galaxy group scale.  Here, we suggest the odds of the bound violations of the turn-around radii of the 
galaxy groups as a new complementary diagnostics. In the standard $\Lambda$CDM cosmology based on GR, the averaged 
turn-around radius of the galaxy groups is bounded by a finite upper limit that depends on the amount of $\Lambda$ as well as on the 
masses of the groups \citep{PT14,pavlidou-etal14}.  
A recent numerical study has revealed that in rare occasions the turn-around radii of individual galaxy groups commit the bound  
violations even in the $\Lambda$CDM cosmology \citep{LY16}. Given that the turn-around radii of the galaxy groups reflect how far the 
expanding spacetime resists the gravitational attraction of the groups, we speculate that the presence of MG would produce a substantial 
difference in the odds of the bound violation of the galaxy groups. The main task that we are to perform in the current work is to 
numerically investigate how strong effect the presence of MG has on the odds of the bound-violation of the turn-around radii of the 
galaxy groups. 

This Paper is divided into three sections the contents of which are summarized as follows. In Section \ref{sec:review} we provides a brief 
review of a certain type of two MG models considered to perform our task and describes the sample of the galaxy groups from N-body 
simulations for the standard $\Lambda$CDM cosmology and for two MG models. 
In Section \ref{sec:vr} we present a detailed description of the procedures by which the odds of the bound violation of the tun-around 
radii of the galaxy groups are calculated for each model. In Section \ref{sec:sum} we summarize the results and discuss the advantages 
of using the odds of the bound violations as a complementary probe of gravity. 

\section{DATA AND MODELS : A BRIEF REVIEW}
\label{sec:review}

As a fiducial model of MG whose effect on the odds of the bound violation of the turn-around radius is to be explored, we focus on the 
normal branch Dvali-Gabadadze-Porrati (nDGP) brane world model \citep{nDGP}. Although the nDGP model is not capable of explaining 
the cosmic acceleration without assuming the existence of some form of dark energy in the universe 
\citep[e.g.,see][and references therein]{schmidt09}, it possesses the following two salient features: 
First, in this model the Hubble parameter $H(z)$ can be made identical to that of the standard $\Lambda$CDM model by \citep{schmidt10}. 
Second, the Vainshtein mechanism \citep{vainshtein72} on which the screening process of this model relies is independent of the shape 
of the potential function of the scalar field \citep{MK10,sbisa-etal12,WF15,falck-etal15}. 

The initial conditions of the nDGP model can be specified by determining the values of seven key parameters: 
the spectral index ($n_{s}$), the baryon density parameter ($\Omega_{b}$), the matter density parameter ($\Omega_{m}$), 
the $\Lambda$ energy density parameter ($\Omega_{\Lambda}$), the Hubble constant ($H_{0}$), linear density amplitude ($\sigma_{8}$), 
cross-over scale ($r_{c}$) \cite[see][and references therein]{falck-etal15}. 
The first six represent the same key cosmological parameters as the standard $\Lambda$CDM model requires, while the last one 
$r_{c}$ is an extra parameter introduced by the nDGP model. 

The cross over scale $r_{c}$ appears in the following modification of the linearized Poisson equation as an additional term to quantify the 
effect of enhanced gravity \citep{KS07}:   
\begin{equation}
\label{eqn:poisson}
\nabla^{2}_{\bf x}\Psi({\bf x},t) = 4\pi\,G\,a^{2}(t)\,\bar{\rho}(t)\delta(t)
\left[1+\frac{1}{3}\left(1 + 2H(t)\,r_{c}+\frac{2}{3}\frac{dH}{dt}\frac{r_{c}}{H(t)}\right)^{-1}\right]\, ,
\end{equation}
where $\Psi({\bf x},t)$ in the Newtonian potential, $H(t)$ is the Hubble parameter and $\bar{\rho}(t)$ is the mean mass density of the 
Universe, and $\delta(t)$ is the dimensionless density contrast. The strength of the fifth force decreases as the cross over scale $r_{c}$ 
increases in the nDGP models. The GR would be restored in the high-density region via the Vainshtein mechanism as 
the value of $r_{c}$ increases with the growth of non-linearity \citep{falck-etal15}.  A comprehensive review on the Vainshtein mechanism 
and the nDGP models is presented in \citet{vainshtein_review}.

Three cosmological models are considered for our numerical exploration. The standard GR+$\Lambda$CDM cosmology and two nDGP 
models denoted as DF5 and DF6. From here on, the two abbreviated terms, the GR and the GR+$\Lambda$CDM cosmology, 
are used interchangeably.  The values of the key cosmological parameters adopted for the three models are listed in 
Table \ref{tab:3models}.  As can be read, except for the values of $r_{c}$ and $\sigma_{8}$, the key cosmological parameters are set at 
the same values of the Planck cosmology without massive neutrinos \citep{planck14}. The highest value of $\sigma_{8}$ of the DF5 
model translates into the strongest fifth force, which still meets the observational constraint from the cluster counts 
\citep{schmidt-etal09,lombriser-etal10,falck-etal15}.  The values of $r_{c}$ used for the DF5 and DF6 models are also compatible with the 
recent observational constraints \citep[see Table 1 in][and references therein]{vainshtein_review}. 

The N-body simulations implemented by the adapted ECOSMOG-V code \citep{ecosmog,code_DGP} were run from $z=49$ to $z=0$ on the 
periodic box of volume $(128\,h^{-1}\,{\rm Mpc})^{3}$ with a total of $512^{3}$ DM particles of individual mass 
$M_{\rm par}=1.34\times 10^{9}\,h^{-1}\,M_{\odot}$, producing five different realizations for each model. 
The Rockstar halo finder developed by \citet{rockstar_finder} has been applied to the phase space distributions of DM particles at each 
$z$-snapshot for the identification of the bound DM halos.  In the catalogs of the Rockstar halos are stored such information on the DM 
halos as the comoving positions and velocities of their centers of masses, virial radii and masses, and so on from each realization for 
each model.  The viral radius of each DM halo $r_{vir}$ has been computed as the spherical radius from its center of mass at which the 
relation of $\rho(r_{vir})=200\rho_{\rm crit}$ is satisfied where $\rho_{\rm crit}$ represents the critical mass density of the Universe. 
Accordingly, the virial mass $m_{vir}$ of each halo has been computed as the mass enclosed by the spherical radius $r_{vir}$. 
 
\section{ODDS OF THE BOUND VIOLATIONS IN NDGP MODELS}

\subsection{Bound-Zone Velocity Profiles in nDGP Models}
\label{sec:vr}

Consider a massive group for which prior information on the virial mass and radius ($M_{vir}$ and $r_{vir}$, respectively) is available.  
Its gravity will influence the peculiar motions not only of its satellites located in the infall zone but also of the neighbor galaxies located in 
the bound zone that corresponds to the distance range of $(3-8)r_{vir}$.  It has been found that for the case of the GR+$\Lambda$CDM 
cosmology, the interplay between the gravity and the expanding space molds the bound-zone velocity profile around a massive group 
to have the following power-law shape \citep{falco-etal14}: 
\begin{equation}
\label{eqn:vr}
\frac{v(r)}{V_c} =  - A\left(\frac{r_{vir}}{r}\right)^{-n}\, ,
\end{equation}
where $A$ and $n$ are the amplitude and the slope parameters of the profile, respectively, and $V_{c}\equiv (GM_{vir}/r_{vir})^{1/2}$. 
After \citet{falco-etal14} reported that Equation (\ref{eqn:vr}) with $A\approx 0.8$ and $n\approx 0.42$ fitted well the {\it average} profile 
$v(r)$ numerically obtained from the bound-zone DM particles around the group-size halos with $M_{vir}\sim 10^{14}\,h^{-1}M_{\odot}$ 
at $z=0$,  it has been proven that Equation (\ref{eqn:vr}) with the same best-fit parameters still describes well eve the average bound-zone 
velocity profiles obtained not from the DM particles but from the bound DM halos \citep{lee16}. 

In the DGP model, the presence of MG is more eminent in the bound zone than in the infall zone since the former has lower densities. 
In other words, in the DGP model the bound-zone objects around a massive group react more sensitively to a fifth force produced by the 
unscreened MG than the infall-zone satellites.  We speculate that the fifth force would decrease the slope of the average bound-zone velocity 
profile since the effective gravity of the massive group enhanced by the fifth force is capable of resisting the expanding space at farther distances.
Before quantitatively investigating if the decrement of the slope of the bound-zone velocity profile will be substantial, however, 
it is first necessary to confirm that the bound-zone peculiar velocity profiles for the nDGP models can be also described by the 
same formula, Equation (\ref{eqn:vr}), whose validity was tested only for the case of the $\Lambda$CDM cosmology.

Putting a mass threshold cut $M_{vir,th}=10^{13}\,h^{-1}M_{\odot}$ on the Rockstar halo catalog described in Section \ref{sec:review}, 
we make a sample of the central groups with virial masses $M_{vir}\ge M_{vir,th}$. 
For each central group in the sample, we look for the neighbor Rockstar halos which satisfy two conditions. First, they should  
belong to the bound zone around the central group with their separation distances $r$ lie in the range of $3\le r/r_{vir}\le 8$.  
Second, the numbers of the particles, $N_{\rm p}$, that comprise a halo are equal to or larger than $20$. The bound-zone halos 
composed of less than $20$ dark matter particles are excluded to avoid possible contamination caused by incomplete condensation. 

Let ${\bf v}_{G}$ and ${\bf v}_{b}$ denote the comoving velocities of a central group and a bound-zone halo, respectively, 
and let ${\bf r}$ be the separation vector from the central group to the bound-zone halo. 
We first subtract ${\bf v}_{G}$ from ${\bf v}_{b}$ to obtain the relative peculiar velocity of the bound-zone halo in the rest 
frame of the central group. Then, we perform the dot product between ${\bf v}_{b}-{\bf v}_{G}$ and $\hat{\bf r}\equiv {\bf r}/\vert{\bf r}\vert$ 
to project the relative peculiar velocity of the bound-zone halo onto the radial direction, $\hat{\bf r}$.   
Let $v$ denote the magnitude of the projected relative peculiar velocity, $\vert ({\bf v}_{b}-{\bf v}_{G})\cdot {\bf h}\vert$, and call it the 
bound-zone velocity at the separation distance $r\equiv \vert{\bf r}\vert$. 

Dividing $v$ and $r$ of each bound-zone halo by the circular velocity of its central group $V_{c}$ and the virial radius 
$r_{vir}$, respectively, we express the rescaled bound-zone velocity profile $\tilde{v}(\tilde{r})\equiv v/V_{\rm c}$ as a function of the 
rescaled separation distance $\tilde{r}\equiv r/r_{vir}$. Note that both of $\tilde{v}$ and $\tilde{r}$ are dimensionless. 
We also divide the range of $\tilde{r}$ into several intervals, $[\tilde{r},\tilde{r}+\Delta\tilde{r}]$, 
each of which has the same length $\Delta\tilde{r}$ and record the numbers of the bound-zone halos, $N_{b}$,  whose 
values of the rescaled distances $\tilde{r}$ belong to each interval.  The mean bound-zone velocity at each 
$\tilde{r}$-interval is computed by taking the average over those $N_{b}$ halos. 

Let $N_{b,k}$ and $N_{T,k}$ denote the numbers of the bound-zone halos around each central group and the number of 
the central groups, respectively, in the $k$th realization of each model. Let also $\tilde{v}_{ij,k}(\tilde{r})$ denote the bound-zone 
velocity of the $i$th halo whose separation distance lies in the range of $[\tilde{r},\tilde{r}+\Delta\tilde{r}]$ around the $j$th central 
group in the $k$th realization of each model. 
The bound-zone velocity profile in the $k$th realization, $\tilde{v}_{k}(\tilde{r})$, can be computed by taking the average first over 
the $N_{\rm b, k}$ bound-zone halos and then over the $N_{\rm T,k}$ central groups as
\begin{equation}
\tilde{v}_{k}(\tilde{r}) = \frac{1}{N_{\rm T,k}} \sum_{j=1}^{N_{\rm T}} 
\left(\frac{1}{N_{\rm b,k}} \sum_{i=1}^{N_{\rm b}} \tilde{v}_{ij,k}(\tilde{r})\right)\, .
\end{equation}

Finally, the bound-zone velocity profile $\tilde{v}(\tilde{r})$ for each model can be obtained by taking the average over the five 
realizations. The errors can be also computed by using the bootstrap technique.  Let $\{\tilde{v}_{k}(\tilde{r})\}_{k=1}^{5}$ denote the 
original sample of the bound-zone velocity profiles from the five realizations. 
From this sample, we draw five bound-zone velocity profiles with repetition allowed to create a boostrap resample. 
We create $1000$ Boostrap resamples and calculate the errors associated with the measurement of the average bound-zone velocity 
profile as the one standard deviation scatter around the mean value: 
\begin{equation}
\label{eqn:boostrap_err}
\sigma^{2}_{v}(\tilde{r}) = \frac{1}{1000}\sum_{\alpha=1}^{1000}
\left[\left(\frac{1}{5}\sum_{k=1}^{5}\tilde{v}_{k}^{\alpha}(\tilde{r})\right) - 
\left(\frac{1}{5}\sum_{k=1}^{5}\tilde{v}_{k}(\tilde{r})\right)\right]^{2}\, ,
\end{equation}
where $\tilde{v}_{k}(\tilde{r})$ represents the average bound-zone velocity profile from the original sample of the $k$th realization and 
$\tilde{v}_{k}^{\alpha}(\tilde{r})$ is from the $\alpha$th Bootstrap resample of the $k$th realization.

\begin{figure}[tb]
\includegraphics[scale=1.0]{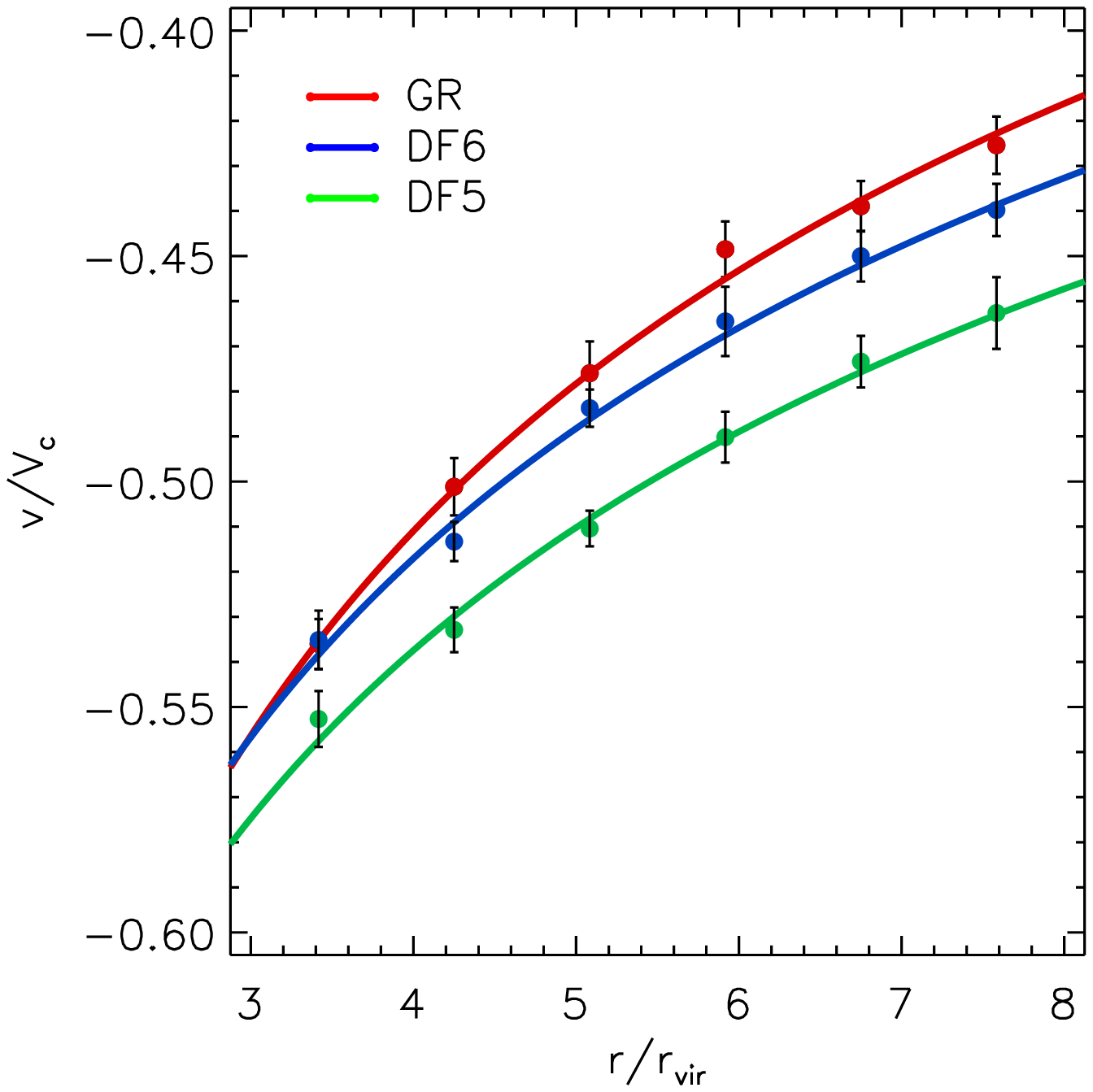}
\caption{Average bound velocity profiles in the bound zone of the central groups with virial mass $M_{\rm v}\ge 10^{13}\,h^{-1}\,M_{\odot}$ 
at $z=0$ for three different models (GR, DF6 and DF5 as red, blue and green colors, respectively).  The filled circles correspond to the 
numerical results while the solid lines are the analytic model, Equation (\ref{eqn:vr}), with the best-fit parameters.}
\label{fig:vr}
\end{figure}

Figure \ref{fig:vr} plots the bound-zone peculiar velocity profile averaged over five realizations as filled circles with the Boostrap errors.  
There is a substantial difference in $\tilde{v}$ between the GR and the DF5 models. The latter has a lower slope than 
the former, as speculated. Furthermore, the magnitude of $\tilde{v}$ in the DF5 model is larger in the whole bound-zone range than 
in the GR model.  Fitting the numerically obtained profile $\tilde{v}(\tilde{r})$ of each model to Equation (\ref{eqn:vr}) by employing the 
maximum likelihood method, we search for the values of $n$ and $A$ which minimize the following $\chi^{2}$.
\begin{equation}
\label{eqn:chi2}
\chi^{2} = \sum_{i=1}^{N_{\rm r}}\frac{\left[\tilde{v}(\tilde{r}_{i})-\tilde{v}^{\rm the}(\tilde{r}_{i}|n_{\rm v},\beta)\right]^{2}}
{\sigma^{2}(\tilde{r}_{i})}\, ,
\end{equation}
where $\tilde{r}_{i}$ denotes the $i$th interval of $\tilde{r}$ and $\tilde{v}^{\rm the}$ represents the theoretical prediction of 
Equation (\ref{eqn:vr}). To determine the uncertainties associated with the determination of $n$ and $A$, we first determine the joint 
probability $p(A,n)=p\left[-\chi^{2}(a,b)/2\right]$.   Using the probability density functions $p(n)$ and $p(A)$ calculated as 
$p(A)=\int\, dn\,p(A,n)$ and $p(n)=\int\, dA\,p(A,n)$, respectively, we also determine the marginalized errors, 
$\sigma_{n}$ and $\sigma_{A}$. 

\begin{figure}
\begin{center}
\plotone{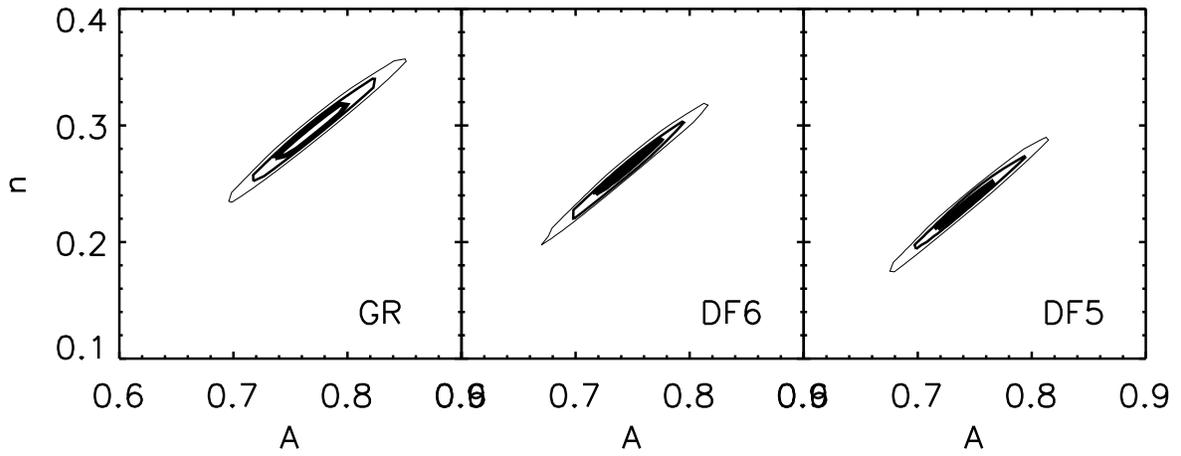}
\caption{$68\%$, $95\%$ and $99\%$ confidence regions (thick, thin and the thinnest solid lines, respectively) 
in the $A$-$n$ plane determined by using the standard Maximum-Likelihood method for three different models. }
\label{fig:cont}
\end{center}
\end{figure}

Figure \ref{fig:cont} shows the boundaries of three different regions in the space spanned by $A$ and $n$ that enclose those points over 
which the integration of $P(A,n)$ becomes $0.68$, $0.95$ and $0.99$ as the thickest, thick and thinnest lines, respectively, for each 
model. The third and fourth columns of Table \ref{tab:3models} display the best-fit values of $A$ and $n$ with the associated errors 
($\sigma_{A}$ and $\sigma_{n})$ for each model.  The solid lines in Figure \ref{fig:vr} correspond to the analytic model whose characteristic 
parameters are set at the best-fit values listed in Table \ref{tab:3models}. 
As can be read, the power-law slope $n$ has the largest (smallest) value for the GR (DF5) case and the difference in the value of $n$ 
between the two models is statistically significant. 
Due to the non-vanishing fifth force, the bound-zone velocity profile in the DF5 model decreases less rapidly with the distance than 
in the GR model. For the case of the DF6 model, the fifth force is not strong enough to produce any statistically significant difference in 
$\tilde{v}(\tilde{r})$ from the GR case, as expected.

The results shown in Figure \ref{fig:vr} have one important implication. The peculiar velocity profiles for the nDGP models are still well fitted 
by Equation (\ref{eqn:vr}) that was empirically derived by \citet{falco-etal14} from N-body simulations for the GR+$\Lambda$CDM cosmology. 
Having no analytic framework within which the bound-zone velocity profile can be derived from the first principles for nDGP models, we have assumed 
that no matter what background cosmology is used, the peculiar velocity of a bound-zone galaxy may be proportional to some power of the separation 
distance and thus that the same functional form of Equation (\ref{eqn:vr}) can still describe the average peculiar velocity profile even for the nDGP 
models.  This assumption is justified by the very fact that the current work has found a good agreement with Equation (2) with the best-fit parameters 
and the numerical results obtained from the N-body simulations for the two nDGP models. 

\begin{figure}[tb]
\includegraphics[scale=1.0]{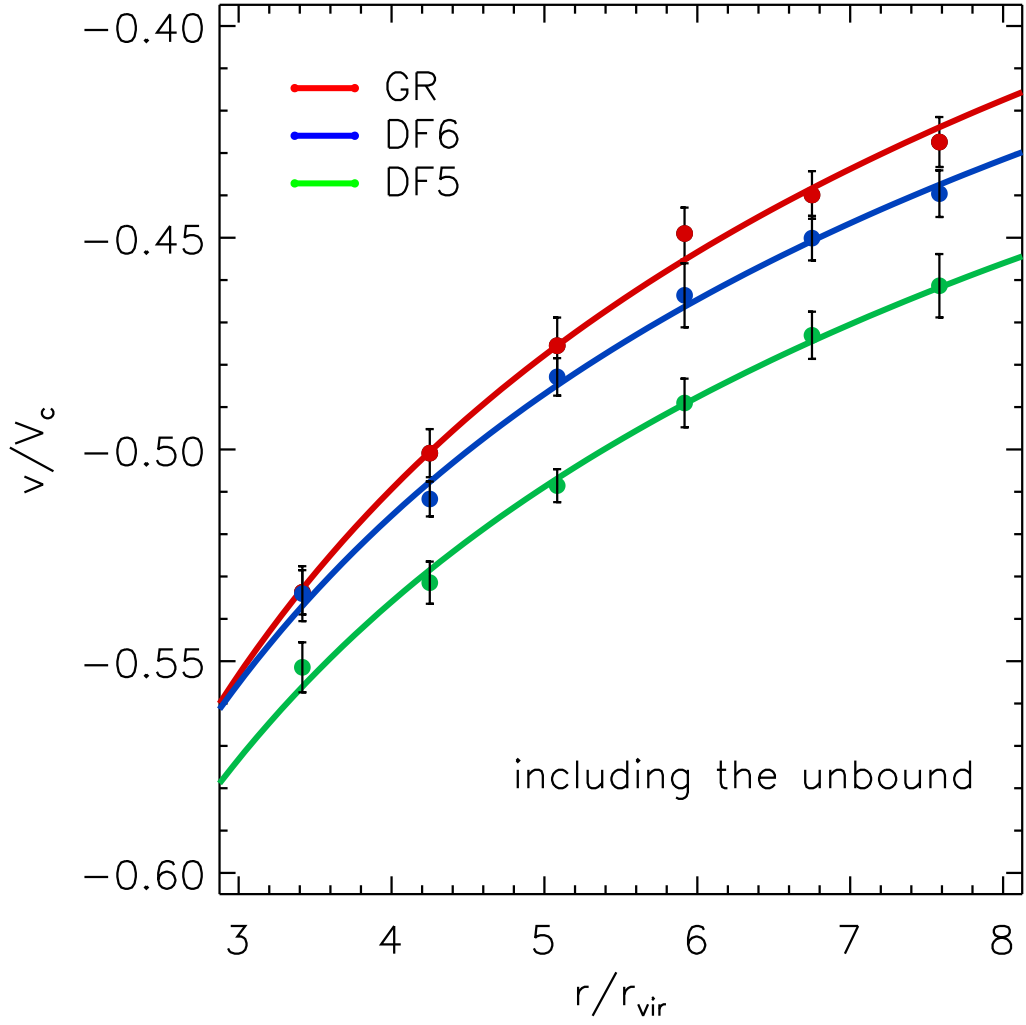}
\caption{Same as Figure \ref{fig:vr} but for the case that the mass of each central group is determined without excluding 
the unbound particles within its virial radius.}
\label{fig:vr_unb}
\end{figure}

Before estimating the turn-around radii by using $\tilde{v}(\tilde{r})$, it may be worth addressing one crucial issue. The Rockstar algorithm 
counts only the bound particles to calculate the virial mass of a halo,  adopting the same definition of the {\it boundedness}  regardless of the 
background cosmology.  However, any departure of the gravitational law from GR may change the concept of {\it being gravitationally bound}, 
the mass of a halo computed by the Rockstar finder may not be the true virial mass, which in turn may change the shape of the peculiar 
velocity profile in the nDGP models. To address this issue, we include the unbound particles within the virial radius of each central group to compute  
its virial mass. Then, repeating the same procedure, we redetermine $\tilde{v}(\tilde{r})$ for the three models, which are plotted in Figure \ref{fig:vr_unb}. 
As can be seen, there is almost no change between the results displayed in Figures \ref{fig:vr} and \ref{fig:vr_unb}, which implies that 
the difference in the definition of the boundness between GR and nDGP models is unlikely to have a significant effect on our estimates of the 
turn-around radii of the central groups from the average bound-zone peculiar velocity profiles.  

\subsection{Turn-Around Radii of the Central Groups in nDGP Models}
\label{sec:rt}

The GR+$\Lambda$CDM cosmology puts an upper bound, $r_{t,u}$, on the average turn-around radius of a galaxy group with mass $M_{vir}$ 
\citep{PT14}:
\begin{equation}
\label{eqn:rt_u}
r_{t,u} = \left(\frac{3M_{vir}G}{\Lambda C^{2}}\right)^{1/3}\, .
\end{equation}
This upper bound, however, limits the {\it expectation value} of the turn-around radius but not the individual values of $r_{t}$ because the 
event of a turning around is a generically random process \citep{PT14}. In other words, the turn-around radii of individual galaxy groups can 
have values larger than $r_{t,u}$ although the occurrences of such bound violations are quite rare \citep{LY16}. 

Strictly speaking, the turn-around radius $r_{t}$ of a galaxy group is an attribute that it acquires at the end of its proto-group stage. 
The optimal way to estimate the turn-around radius of a central group in a N-body experiment is to track the trajectories of the component 
DM particles back to the proto-group regime and then to find the location at which the mean peculiar velocity equals the Hubble speed.  
This optimal routine, however, is simply impractical and thus not applicable to real data. 
Recently, \citet{lee-etal15} formulated a less optimal but much more practical routine that makes it possible to estimate the turn-around 
radius of a galaxy group from the direct observables. This routine counts on Equation (\ref{eqn:vr}) to find $r_{t}$ at which 
the following equation holds true: 
\begin{equation}
\label{eqn:rt}
\frac{H_{\rm 0}r_{t}(M_{vir})}{V_{c}} = A\,\left[\frac{r_{t}(M_{vir})}{r_{vir}}\right]^{-n}\, .
\end{equation}

In Section \ref{sec:vr} we determine the best-fit values of $A$ and $n$ for the mean bound-zone velocity profile averaged over the central 
groups. To use Equation (\ref{eqn:rt}) to estimate $r_{t}$ of each central group, however,  it is necessary to determine the values of $A$ and 
$n$ by separately fitting the individual bound-zone velocity profile around each central group to Equation (\ref{eqn:vr}).  
Let $\tilde{v}_{j,k}(\tilde{r})$ be the bound zone velocity profile of the $j$th central group in the $k$th realization of each model. 
Replacing $\tilde{v}_{k}$ in Equation (\ref{eqn:chi2}) by $\tilde{v}_{j,k}(\tilde{r})$ and minimizing $\chi^{2}$, we determine the 
best-fit values of $A$ and $n$ and put them into Equation (\ref{eqn:rt}) to estimate the turn-around radius of the $j$th central group 
in the $k$th realization, say $r_{t,j,k}(M_{vir})$. 

\begin{figure}[tb]
\includegraphics[scale=1.0]{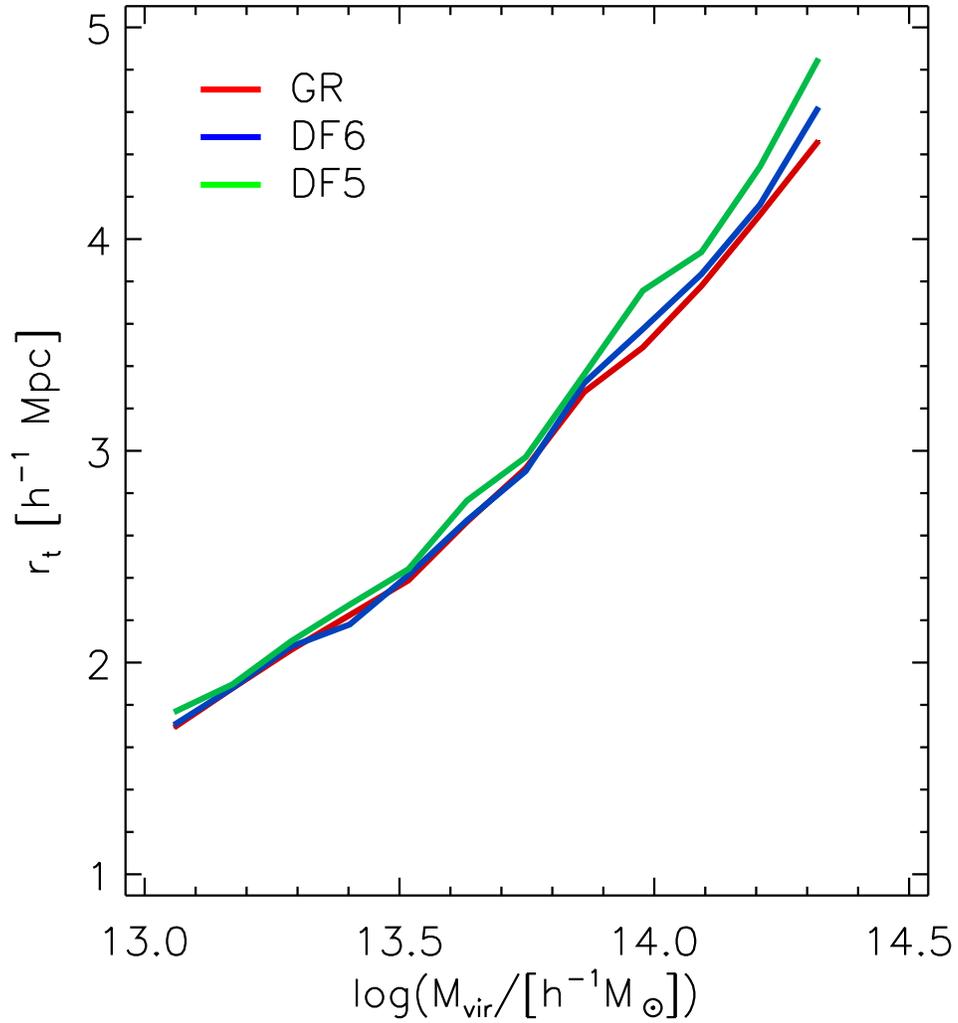}
\caption{Turn-around radii of the central groups versus the logarithmic masses for three different models 
(GR, DF6 and DF5 as red, blue and green colors, respectively). }
\label{fig:rt_all}
\end{figure}

Dividing the range of the logarithmic masses of the central groups, $m_{vir}\equiv \log(M_{vir} h\,M^{-1}_{\odot})$, into several intervals 
each of which has the same length $\Delta m_{vir}$, we calculate the mean turn-around radius, $r_{t,k}(M_{vir})$, by taking the average 
of $r_{t,j,k}(M_{vir})$ over the central groups whose logarithmic masses lie in a given interval of $[m_{vir}, m_{vir}+\Delta m_{vir}]$. 
The mean turn-around radius, $r_{t}$, of a central group at each $m_{vir}$-interval is now evaluated as taking the average of 
$r_{t,k}$ over the five realizations for each model. Figure \ref{fig:rt_all} plots $r_{t}$ versus $m_{vir}$ for the GR, DF6 and DF5 
models as red, blue and green solid lines, respectively. As can be seen, there is a notable difference in $r_{t}(m_{vir})$ between the GR 
and the DF5 models. To see whether or not this difference is statistically significant, we calculate the Bootstrap errors 
$\sigma_{r}$ in the estimation of $r_{t}$ by generating $1000$ Bootstrap resamples as done in Section \ref{sec:vr}. 

\begin{figure}[tb]
\includegraphics[scale=1.0]{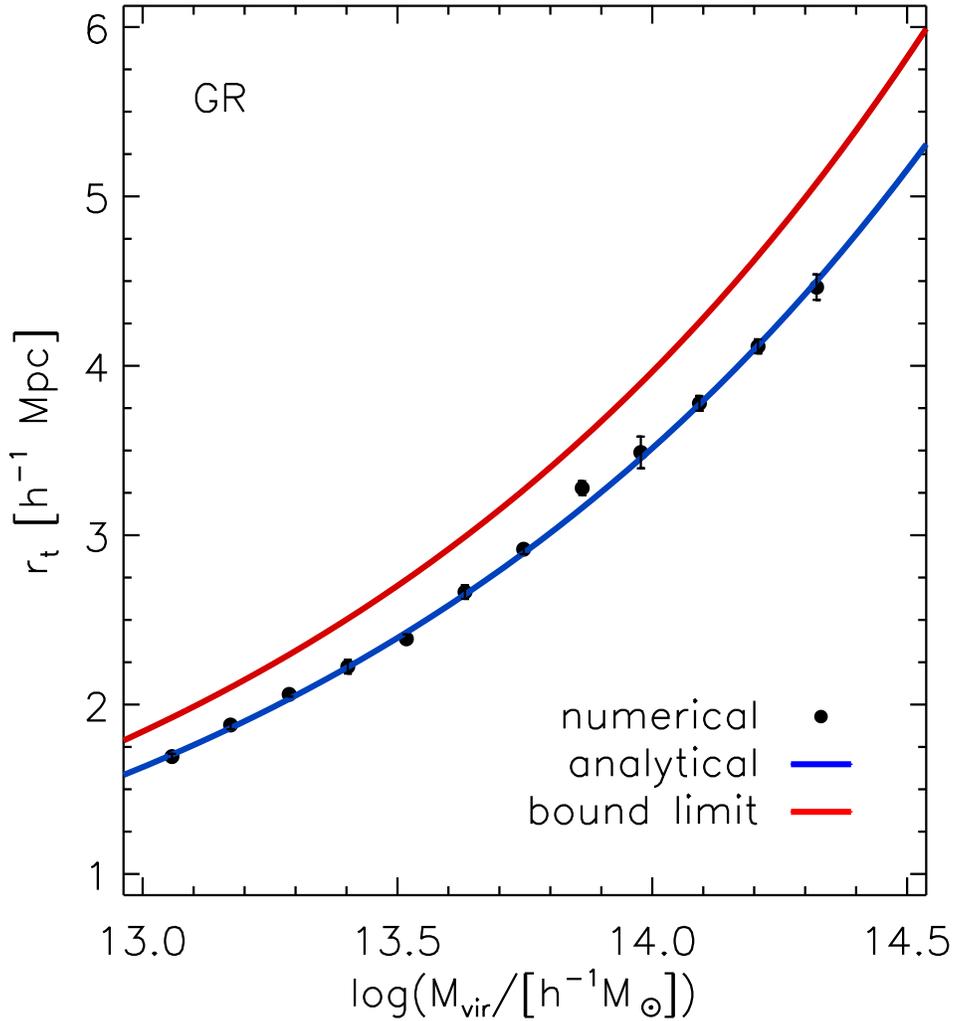}
\caption{Turn-around radii of the central groups as a function of their masses in the logarithmic scales for the GR model. 
The black filled circles with Boostrap errors represent the numerical results while the blue solid line represents the analytic 
results evaluated by using Equation (\ref{eqn:rt}). The red solid line is the upper bound limit, Equation (\ref{eqn:rt_u}) 
predicted by the GR+$\Lambda$CDM \citep{PT14}.}
\label{fig:rt_gr}
\end{figure}

\begin{figure}[tb]
\includegraphics[scale=1.0]{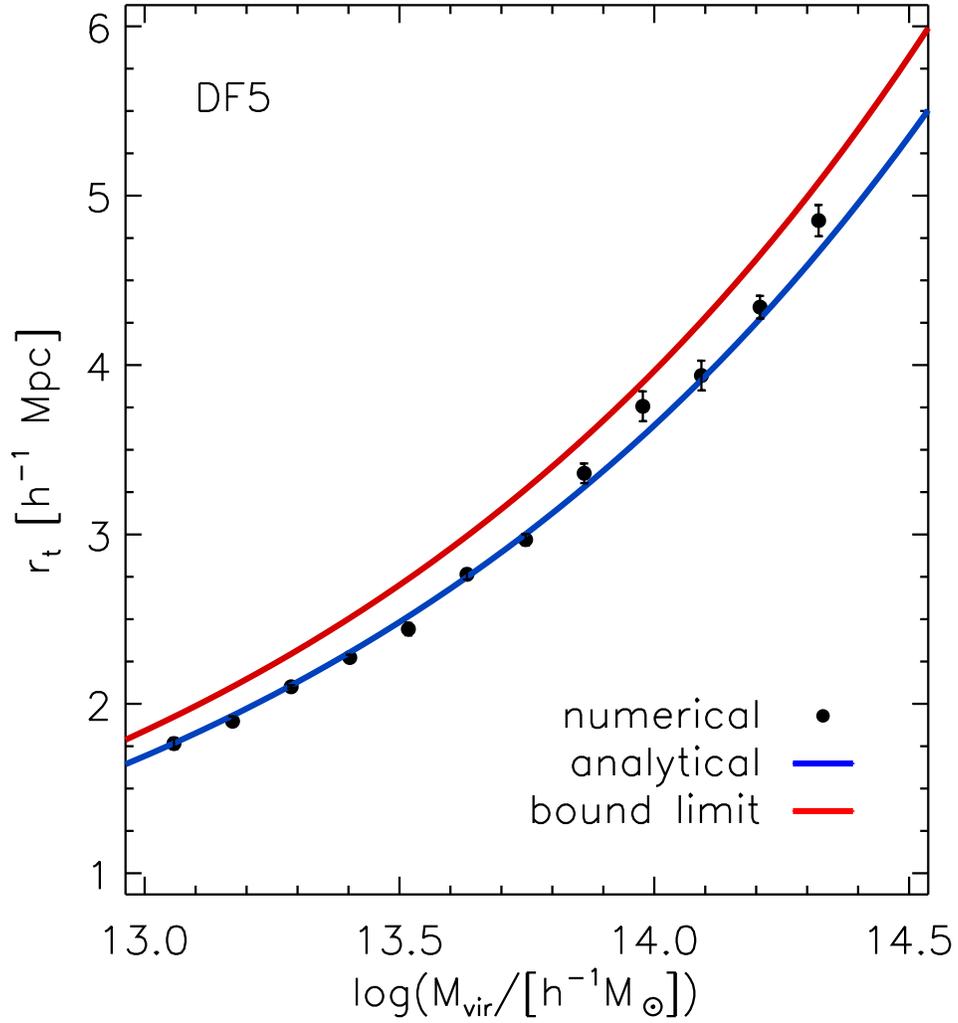}
\caption{Same as Figure \ref{fig:rt_gr} but for the DF5 model.}
\label{fig:rt_df5}
\end{figure}

Figure \ref{fig:rt_gr} shows $r_{t}(m_{vir})$ as filled circles with the Bootstrap errors for the GR case. The red solid line represents the 
upper bound limit $r_{t, u}$ given in Equation (\ref{eqn:rt_u}).  As can be seen, the mean turn-around radius is lower than the upper limit in the 
whole range of $m_{vir}$. The blue solid line is obtained by putting the global average values of $A$ and $n$ listed in Table \ref{tab:3models} 
into Equation (\ref{eqn:rt}) and solving it for $r_{t}$.  Note that the blue solid line is in good agreement with the filled circles in the entire 
range of $m_{vir}$, which is consistent with the claim of \citet{LY16} that the average turn-around radii can be computed by using the 
average bound-zone velocity profile with the two parameters set at the universal best-fit values. 
Figure \ref{fig:rt_df5} plot the same as Figure \ref{fig:rt_gr} but for the case of DF5 model. As can be seen, similar to the GR case, 
the DF5 model yields the average turn-around radii lower than the upper limit $r_{t,u}$ in the entire range of $m_{vir}$. Note, however, 
that the gap between the average turn-around radii and the upper bound limit is narrower in the DF5 model than in the GR case. 

Now, we are ready to calculate the odds of the bound violations, for which we exclude those central groups whose logarithmic masses 
exceed $14.5$. Given that the DF5 model produces more massive groups in the highest mass section than the other two models, it might 
cause a bias in the calculation of the odds of the bound-violations if those central groups with $m_{vir}\ge 14.5$ were not excluded.
Let $r_{t, k}$ denote the turn-around radii of the central objects with mass $m_{vir}$ in the $k$th realization of each model. 
Define $\eta$ as the ratio of $r_{t,k}$ to the upper bound $r_{t, u}$ as $\eta\equiv r_{t,k}/r_{t, u}$. 
Dividing the whole range of $\eta$ into several small bins each of which has the same length $\Delta\eta$,  and counting the 
numbers of the central groups whose ratios belong to each bin, $[\eta, \eta+\Delta\eta]$, we first compute the probability density 
function, $p_{k}(\eta)$,  and then integrate $p_{k}(\eta)$ over $\eta$, to derive the cumulative probability distribution, 
$P_{k}(\ge \eta)$ from the $k$th realization of each model. 
The first five columns of Table \ref{tab:numb0} list the values of $P_{k}(\ge\eta)$ at $\eta=1$ for the three models. These values equals 
the ratio of the bound violating central groups to the total number of the central groups in each realization. 

\begin{figure}[tb]
\includegraphics[scale=1.0]{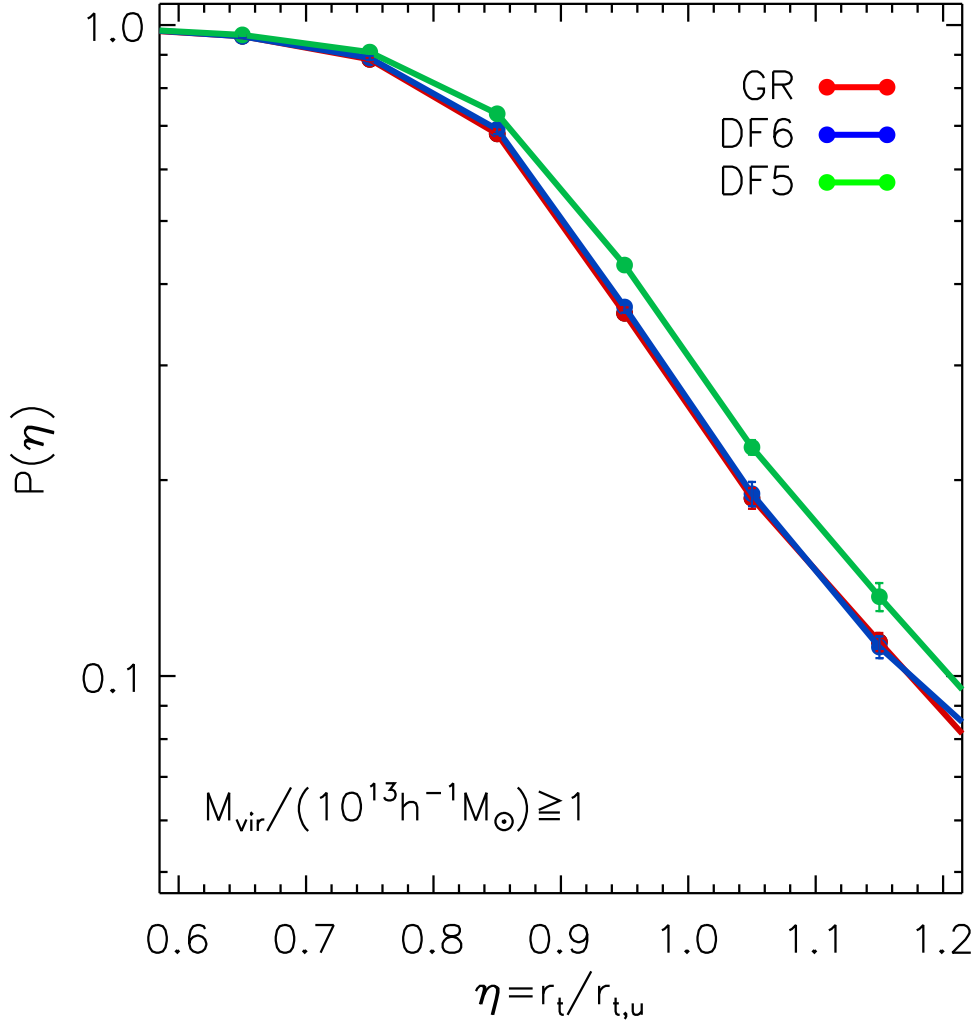}
\caption{Cumulative probabilities that the ratio $\eta\equiv r_{t}/r_{t,u}$ with bootstrap errors for three different models 
(GR, DF6 and DF5 as red, blue and green colors, respectively). The central groups with masses in the range of 
$10^{13}\le M/(h^{-1}M_{\odot})\le 10^{14.5}$ is considered.}
\label{fig:cpro}
\end{figure}

Finally, we take the average of the cumulative probability functions over the five realizations for each model as
 \begin{equation}
 \label{eqn:eta}
 \bar{P}(\ge\eta) = \frac{1}{5}\sum_{k=1}^{5}P_{k}(\ge\eta)\, .
 \end{equation}
Figure \ref{fig:cpro} plots $\bar{P}(\ge \eta)$ versus $\eta$ with the Bootstrap errors for the three models. As can be seen, the DF5 (GR) 
model yields the highest (lowest) values of $\bar{P}(\ge \eta)$ in the range of $\eta\ge 0.8$. In other words, in the DF5 model the bound 
violation occurs relatively more frequently than in the GR model. Meanwhile, no significant difference is found between the GR and the 
DF6 models, as expected. 

The sixth column of Table \ref{tab:numb0} lists the average odds of the bound violations, $\bar{P}(\ge 1)$, for the three models. 
In the GR model the odds of the bound violations is $0.188\pm 0.007$ while in the DF5 model it is $0.225\pm 0.006$. As speculated, 
in the DF5 model due to the fifth force the bound violations occur relatively more frequently. 
The signal to noise ratio for the difference in the odds of the bound-violations between the two models is found to be as high as $4.2$.
This result indicates that the odds of the bound-violations of the turn-around radii of the central groups can be a powerful indicator 
of the presence of MG. 

\begin{figure}[tb]
\includegraphics[scale=1.0]{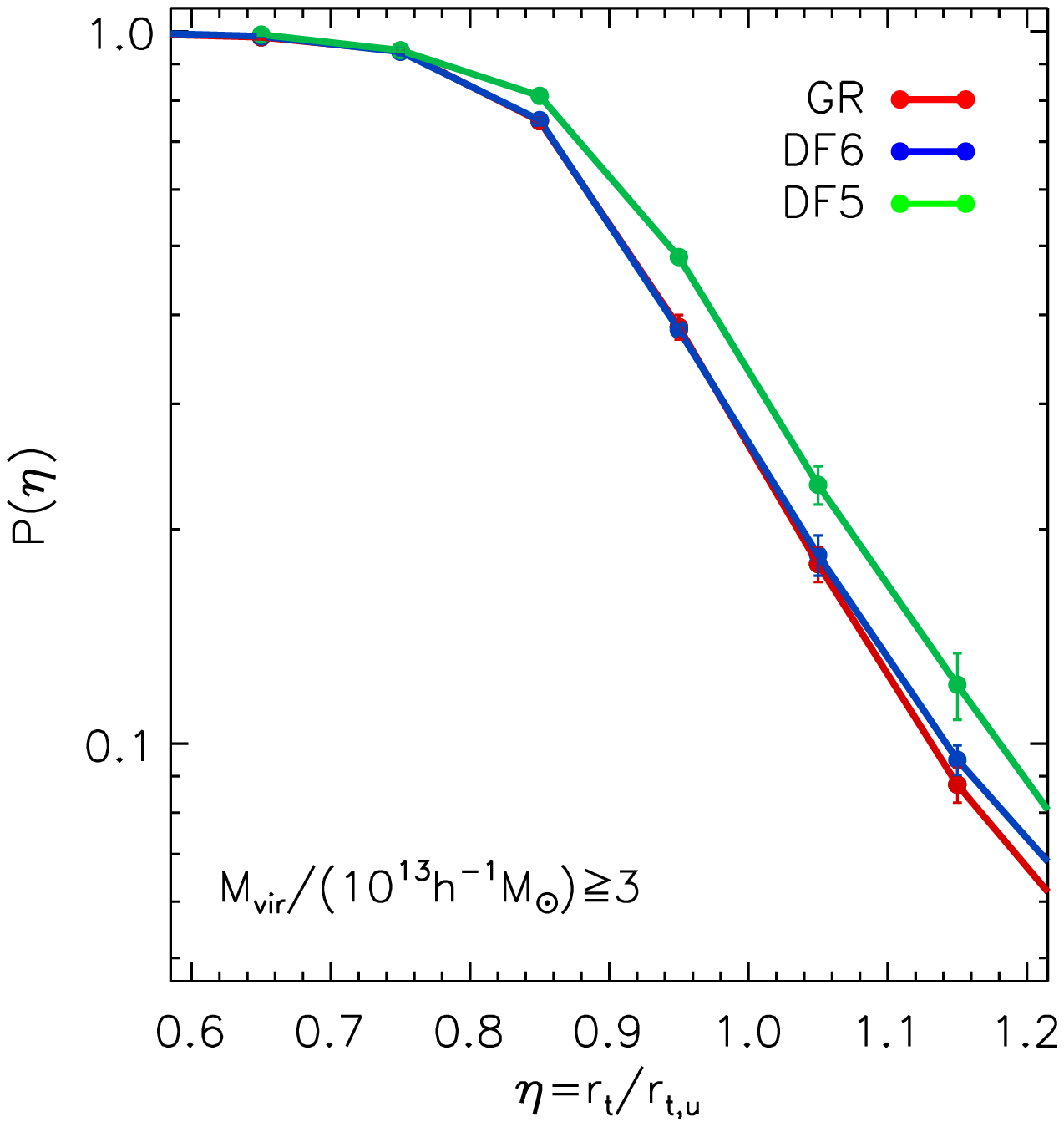}
\caption{Same as Figure \ref{fig:cpro} but with the central groups with masses in the range of 
$3\times 10^{13}\le M/(h^{-1}M_{\odot})\le 10^{14.5}$.}
\label{fig:cpro3}
\end{figure}

\begin{figure}[tb]
\includegraphics[scale=1.0]{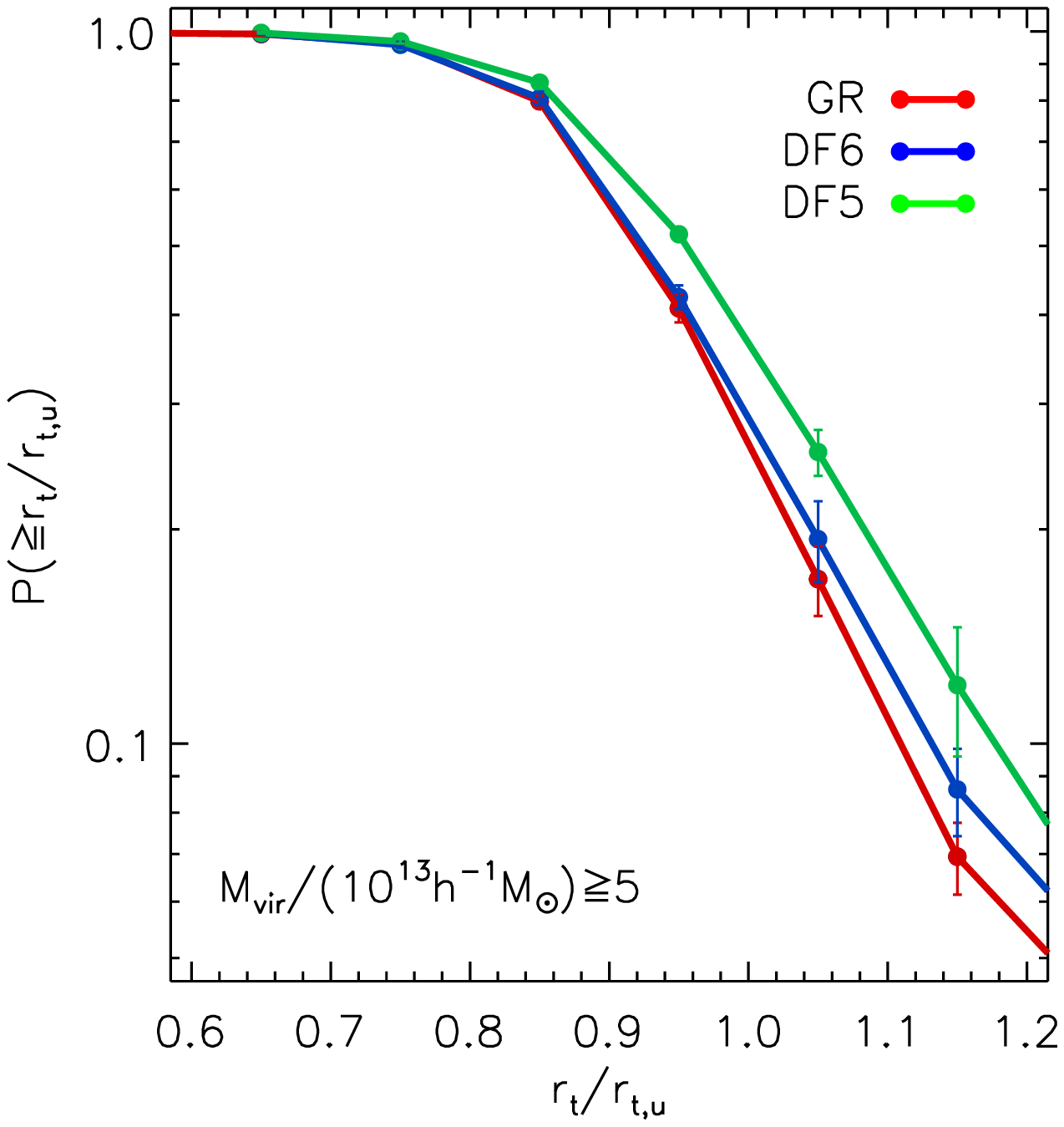}
\caption{Same as Figure \ref{fig:cpro} but with the central groups with masses in the range of 
$5\times 10^{13}\le M/(h^{-1}M_{\odot})\le 10^{14.5}$.}
\label{fig:cpro5}
\end{figure}

To examine if the odds of the bound violations depend on the mass threshold of the central groups, we increase the value of $M_{vir,th}$ 
and rederived $P(\ge\eta)$ by repeating the whole process described in the above. Figure \ref{fig:cpro3} and \ref{fig:cpro5} plot the same as 
Figure \ref{fig:cpro} but for the cases of $M_{vir,th}=3\times 10^{13}\,h^{-1}M_{\odot}$ and $M_{vir,th}=5\times 10^{13}\,h^{-1}M_{\odot}$, 
respectively. As can be seen, no substantial change is made by increasing the values of $M_{vir,th}$. Due to the lower numbers of the 
central groups for these two cases, however, the cumulative probabilities have larger errors and the odds of the bound violations have lower 
signal-to-noise ratios of $\sim 3$.

\section{SUMMARY AND DISCUSSION}
\label{sec:sum}

The turn-around radius of a massive group can be determined as the distance at which the average velocity of its bound-zone galaxies 
becomes equal to the Hubble speed \citep{lee-etal15}. The average bound-zone velocity profile was shown by several numerical 
experiments to follow a power-law scaling \citep{cuesta-etal08,falco-etal14,lee16,LY16}.  Since the slope of the profile reflects how far and 
strongly the gravitational attraction of a massive group can resist the accelerating Hubble flow, any departure of the real gravitational law 
from the GR would change the slope of the bound-zone velocity profile and accordingly  the turn-around radius of the group. The question 
is whether or not the change would be substantial. In the current work, we have conducted a numerical analysis to find a quantitative 
answer to this question. 
With the help of the N-body simulations performed for two nDGP models (DF5 and DF6) as well as for the standard $\Lambda$CDM 
+ GR cosmology,  we have measured the slopes of the average bound-zone velocity profiles around the massive groups with masses 
in the range of $10^{13}\le M/(h^{-1}M_{\odot})\le 10^{14.5}$ and found that the DF5 (GR) model yields the lowest (highest) slope. 

Our explanation for this result is that the gravitational attraction enhanced by the fifth force resists the Hubble flow at larger distances, 
resulting in a milder decrease of the average bound-zone velocity profile with the distance in the DF5 model. 
We have also estimated the turn-around radii of the massive groups from the individual bound-zone velocity profiles for each model 
and calculated the odds of the bound-violations by counting the numbers of the massive groups whose estimated turn-around radii 
exceed the bound limit predicted by the GR+$\Lambda$CDM cosmology.  
A $4\sigma$ signal of difference has been found in the odds of the bound-violations between the GR and the DF5 models.  Given that 
the cross-over scale $H_{0}r_{c}/c=1.2$ used for the DF5 model is compatible with the current constraints from the large-scale structure 
observations $H_{0}r_{c}/c\ge 1.0$ \citep[see Table 1 in][and references therein]{vainshtein_review},  
we suggest that it may be in principle possible to use the odds of the bound-violations as a complementary diagnostics to test locally the 
nature of gravity. 

This new diagnostics has a good advantage over the conventional ones. As explained well in \citet{PT14}, the estimate of the 
turn-around radius is not affected by complicated baryonic processes unlike the other local probes of gravity such as the redshift distortion 
effect, N-point correlation functions, and cluster abundances. In other words, a direct comparison between the observational result and 
the theoretical prediction for the odds of the bound violations can be made without taking into account the non-gravitational effects of baryon 
physics, which are hard to model theoretically due to their highly nonlinear nature. Furthermore, since the turn-around radius is a uniquely defined 
quasi-linear quantity,  there is no ambiguity involved with the details of the way that it is estimated.

Moreover, \citet{lee16} has proven that the best-fit values of the amplitude and slope parameters, $A$ and $n$, of Equation (\ref{eqn:vr}) are 
indeed insensitive to the variation of the key cosmological parameters within the GR+$\Lambda$CDM cosmology, which implies that this 
new diagnostics is robust against the changes of the initial conditions in the standard picture. In other words, if the new diagnostics finds a 
tension with the prediction of the GR+$\Lambda$CDM cosmology, then it is quite unlikely that the tension can be alleviated by varying the key 
cosmological parameters or the baryon physics within the GR+$\Lambda$CDM cosmology. 

Another promising aspect of this new diagnostics is that it has a power to distinguish among the MG models with different screening 
mechanisms. Recently, \citet{LY16} have shown that the odds of the bound violations become larger if the odds are calculated from the 
bound-zone velocity profiles constructed along the filaments.  In fact, in order to apply the routine of \citet{lee-etal15} to real observational 
data for the estimate of the turn-around radius of a galaxy group, it is a prerequisite to find a filamentary structure in the bound zone and 
to construct the bound-zone velocity profile along the filament \citep{falco-etal14,lee-etal15} since the anisotropic distribution of the 
bound-zone galaxies along the filaments allows us to construct the bound-zone velocity profile without measuring accurately the peculiar 
velocities \citep[see][for details]{falco-etal14}. 

According to \citet{falck-etal15}, the fifth force in a filament remains intact by the Vainshtein screening mechanism no matter how overdense 
the filamentary environment is while it can be completely shielded in the dense filamentary environment by the other screening mechanisms like 
the Chameleon \citep[see also][]{falck-etal14}.  When the turn-around radii of the galaxy groups are determined from the bound-zone velocity 
profile constructed along the filaments via the routine of \citet{lee-etal15},  the odds of the bound-violations would become different from the 
predictions of the GR+$\Lambda$CDM case only for the case of the Vainshtein mechanism.  To quantitatively verify this speculation, it will 
require N-body simulations with high resolution performed on a large volume for various MG models with different screening mechanisms, 
so that the dense filaments can be identified around massive groups from the simulation datasets.  

It is worth mentioning here that we have not assessed the practical feasibility of this new local diagnostics as a complementary test of gravity, 
which is beyond the scope of this paper.  Before applying this new diagnostics to real observations, however, what has to precede is to quantitatively 
examine how strongly the odds of the bound violation would be affected by observational uncertainties.  Especially, the systematics associated 
with the measurements of the masses of the galaxy groups should be thoroughly examined since the odds of the bound violations is strongly affected 
by the degree of the accuracy with which the masses of the central groups are estimated. Our future project will take off along this direction. 

\acknowledgments

We thank our anonymous referee for very helpful suggestions. 
J.L. acknowledges the support of the Basic Science Research Program through the NRF of Korea funded by the Ministry of Education 
(NO. 2016R1D1A1A09918491).  J.L. was also partially supported by a research grant from the National Research Foundation (NRF) of Korea 
to the Center for Galaxy Evolution Research (NO. 2010-0027910). BL acknowledges supports by STFC Consolidated 
Grant Nos. ST/L00075X/1 and RF040365. The simulations described in this study used the DiRAC Data Centric system at Durham University, operated 
by the Institute for Computational Cosmology on behalf of STFC DiRAC HPC Facility (www.dirac.ac.uk). This equipment was funded by 
BIS National E-Infrastructure Capital grant ST/K00042X/1, STFC Capital grant ST/H008519/1, and STFC DiRAC Operations grant ST/K003267/1 and 
Durham University. DiRAC is part of the National E-Infrastructure.

\clearpage
\clearpage
\clearpage
\clearpage
\clearpage
\clearpage
\clearpage
\clearpage
\clearpage
\clearpage
\begin{deluxetable}{ccccc}
\tablewidth{0pt}
\setlength{\tabcolsep}{5mm}
\tablecaption{Cross-over scales, linear power spectrum amplitude,  and best-fit parameters of the bound-zone 
velocity profiles for three models.}
\tablehead{models &  $H_{0}r_{c}/c$ & $\sigma_{8}$ & $A$ & $n$}
\startdata
GR  & $\infty$ & $0.83$ & $0.77\pm 0.02$ & $0.30\pm 0.02$ \\
DF6 & $5.65$ & $0.84$ & $0.74\pm 0.02$ & $0.26\pm 0.02$  \\
DF5 & $1.20$ & $0.85$ & $0.74\pm 0.02$ & $0.23\pm 0.02$  \\
\enddata
\label{tab:3models}
\end{deluxetable}
\clearpage
\begin{deluxetable}{ccccccc}
\tablewidth{0pt}
\setlength{\tabcolsep}{5mm}
\tablecaption{Odds of the bound violations for three models.}
\tablehead
{models &  $P_{1}(\ge 1)$ & $P_{2}(\ge 1)$ & $P_{3}(\ge 1)$ & $P_{4}(\ge 1)$ & $P_{5}(\ge 1)$ & $\bar{P}(\ge 1)$}
\startdata
GR  & $109/537$ & $108/539$ & $89/554$ & $102/527$ & $95/535$ &$0.188\pm 0.007$ \\
DF6  & $110/534$ & $105/541$ & $88/554$ & $116/548$ & $91/509$ &$0.190\pm 0.008$\\
DF5  & $130/532$ & $128/539$ & $119/544$ & $116/508$ & $108/519$ &$0.225\pm 0.006$\\

\enddata
\label{tab:numb0}
\end{deluxetable}
\end{document}